\documentstyle[twocolumn,prl,aps]{revtex}
\begin{document}
\twocolumn[\hsize\textwidth\columnwidth\hsize\csname @twocolumnfalse\endcsname

\draft
\title{Divergence of the localization length in quantum Hall systems}

\author{Michael Hilke}
\address{D\'epartement de Physique Th\'{e}orique,
      Universit\'e de Gen\`eve,
      24 quai Ernest Ansermet,
      CH - 1211 Gen\`eve 4, Switzerland}
\date{May 23,1995}
\maketitle
\begin{abstract}
The localization properties of a two-dimensional disordered electron gas in a
strong external magnetic field are studied. The impurities are considered to
be located on a square
lattice
with random amplitudes. The concentration of these impurities is low, i.e.,
the average distance
between the impurities
exceeds the magnetic length.
For short-ranged impurity potentials
 we analytically obtain an exponent $\nu=2/3$ for the divergence of
the localization length at the lowest Landau level. The density of states is
also calculated.

\end{abstract}

]

\narrowtext

 One of the important features of Quantum Hall systems is the competition
 between localized
states and extended states. In the absence of a magnetic field all states
are localized in a
two-dimensional
disordered system, whereas with a sufficiently strong magnetic field the
states at the Landau
levels are extended. This behavior can be described by a diverging
localization length $\xi$
at the
Landau levels $E_n$, which scales with energy as $\xi\sim|E-E_n|^{-\nu}$.
 This transition is studied in two types of experiments. The first one,
 pionered
by Wei {\em et al.} \cite{wei} measured the temperature dependent
magneto-resistivity in $InGaAs/InP$
heterostructures and showed that the half-width $\Delta B$, of the
transition regime where
 $\rho_{xx}$ is nonzero, vanishes as
$\Delta B\sim T^\kappa$, with $\kappa=0.42\pm 0.04$ independent of the
Landau level index. $B$
is the applied magnetic field.
Later on Koch {\em et al.} \cite {koch1} measured values of $\kappa$
ranging between $0.36$ and
$0.81$ for $AlGaAs/GaAs$ systems. This exponent $\kappa$, arises from a
competition between
the phase coherence length, which scales
with temperature as
$L_\Phi\sim T^{-p/2}$, and the localization length $\xi$.
 Using scaling arguments \cite{pruisken} these
exponents can be combined to show that the temperature dependence of
$\rho_{xx}$ in the transition
regime is
controlled by the exponent $\kappa=p/2\nu$.
Very recently Wei {\em et al.} \cite {wei2} measured $p$
with an electron heating experiment in a $InGaAs/InP$ system and found
$p\simeq2$. Combining all
this together yields for
the exponent $\nu\simeq2.38$. The value $p=2$ corresponds to the
electron-electron scattering
 value \cite{enz} with no
external magnetic field. The interpretation of the experiments therefore
remains unclear.
The other type of experiment measures the system-size
dependence of the
width in the transition regime of the
$\rho_{xx}$ peak. These measurements were undertaken by Koch {\em et al.}
\cite{koch2} on $AlGaAs/GaAs$
heterojunctions and they found $\nu=2.3\pm0.1$. For $\kappa$ they measured
$\kappa=0.68\pm0.04$. Here
again the interpretation remains to be studied. For, in order to be
meaningful, this experiment
requires the phase coherence
length $L_{\Phi}$ to
 exceed the system length $L$. This would however induce quantum mesoscopic
 conductance fluctuations
which are not observed in these experiments.
 Another point which remains undecided is the importance of
 {\em electron-electron}
correlations in these experiments.

Analytically no general results for the exponent exist \cite{hajdu}.
 But
in the limit of a very smooth
disorder potential the problem can be mapped onto a classical percolation
picture \cite{trugman}
and yields $\nu=4/3$. Mil'nikov and Sokolev \cite{mil} included quantum
tunneling
and obtained
$\nu=7/3$. Using perturbation theory Hikami \cite{hikami} calculated the
inverse participation
number, which yields $\nu\simeq2.4$ when analyzed in terms of
multifractality \cite{Janssen}.
Other results are obtained numerically. We mention here only the more
recent results
\cite{hajdu}.
Two main approaches can be identified.
The first one studies numerically the Thouless-number;
using this method Ando \cite{ando} has shown that the exponent $\nu$ is
independent of the type of
scattering, i.e., long-range or short-range. He moreover obtains
$\nu\simeq2$ for the lowest Landau
level. The other class of
numerical studies uses the finite-size scaling method. Recent results
include the approach of
Huckestein and Kramer \cite{hucke} who introduced a random-matrix model
in which matrix elements
of the Hamiltonian are given by a random complex number resulting in
correct moments up to second order.
Their latest result for the lowest Landau level is $\nu=2.35\pm0.03$.
Another very recent numerical
result has been obtained by Liu and Das Sarma \cite{liu} using a transfer
matrix combined with  the
finite-size scaling method. They studied different impurity concentrations
and potential ranges.
For the lowest Landau level they obtain values for $\nu$ ranging from $1.86$
to $2.32$.
All these numerical studies consider a dense concentration of impurities,
relative to the magnetic
length $l_0=\sqrt{\hbar c/eB}$, where the
localization length diverges only in the limit when the width of the system
tends to infinity.

In this
letter we consider the opposite limit, where the concentration of impurities
is low enough to
maintain at least one extended state at the lowest Landau level. This
extended state appears even
if the system size is
not infinite.
 The impurities we consider are placed on a lattice of constant $d$ and
their amplitudes are taken randomly.
Furthermore we suppose that no Landau level mixing is present as we consider
very high magnetic
fields and we restrict ourselves to the lowest Landau level.
The system is mapped onto a one dimensional disordered discrete
tight-binding equation using as a
basis the Landau functions. Only the nearest neighbors are considered
as relevant, which is a valuable
approximation, as we examine a sufficiently large elementary lattice
of width $a$.
 The resulting one dimensional diluted Anderson model can be solved
 exactly at the lowest Landau
level $E_0$ and yields an infinite localization length. Around this
energy the localization length
$\xi$ can be evaluated and scales as $\xi\sim(E-E_0)^{-2/3}$ for
short-ranged
impurity potentials. This
demonstrates that we are in a different universality class than the
models studied numerically.
In addition, the density of states is found to scale as
$\rho\sim(E-E_0)^{-1/3}$.

We start by considering a two-dimensional plane of size $L_X\times L_Y$.
In this plane we put a rectangle
filled with impurities whose dimension is $N\times M$. The sizes $L_X$
and $L_Y$ are essentially
meant to be infinite, whereas $N$ and $M$ are finite. In particular $M$
stays finite when we take
the limit $N\rightarrow\infty$. Therefore the localization properties are
studied
in the $X$ direction. We start with the usual Hamiltonian with a magnetic
field and a disordered
potential $V(x,y)$. In the Landau gauge, with the vector potential
$\vec{A}=(B y,0,0)$, where $B$ is the perpendicular magnetic
field, the Landau wave-functions $\Psi_{n,k}$ are eigenfunctions of the
Hamiltonian
without disorder and the spectrum is given
by the Landau levels $E_n=\hbar\omega_c(n+1/2)$, where $\omega_c=eB/mc$ is
the cyclotron frequency.
The Landau-wave functions
of the lowest Landau level are
\begin{equation}
\Psi_{0,k}=N_0 e^{ikx}e^{-(y-kl_0^2)^2/2l_0^2},
\end{equation}
where $N_0=1/\sqrt{L_Xl_0\pi^{1/2}}$ and all constants are taken to be one.
As we work in the very high magnetic field limit
we can assume that no Landau level mixing is present, therefore we will
restrict ourselves
to the lowest Landau level, i.e., $E_0=\hbar\omega_c/2$. We write the
eigenfunction of the
Hamiltonian with disorder as
\begin{equation}
\Psi(x,y)=\int_{k=-\infty}^{\infty} \alpha(k)\Psi_{0,k}(x,y) dk.
\end{equation}
We further suppose that we have a crystal of lattice spacing $a$ and
rewrite (2) in this new basis
as
\begin{equation}
\Psi(x,y)=\sum_{l=-\infty}^{\infty}\psi_l |l\rangle,
\end{equation}
where $\alpha(k)=\sum_{l=-\infty}^{\infty}\psi_l e^{-ikla}$ and
\begin{equation}
| l\rangle=\frac{1}{\sqrt{L_Yl_0\pi^{1/2}}}e^{-(x-la)^2/2l_0^2}
e^{-iy(x-la)/l_0^2}.
\end{equation}
The states $|l\rangle$ have the usual orthonormalization property,
i.e.,
$\langle p|l\rangle=\delta_{l,p}$.
Writing the Schr\"odinger equation using (3) gives
the following discrete one-dimensional
 equation
\begin{equation}
\sum_{l=-\infty}^{\infty} D_{p,l}\psi_l=\epsilon\psi_p  ,
\end{equation}
where $\epsilon=E-E_0$ and
$D_{p,l}=\langle p|V(x,y)|l\rangle$.
The potential of the impurities will be considered to be Gaussian,
with their sites
and amplitudes taken randomly.
It can be expressed
as follows:
\begin{equation}
V(x,y)=\sqrt{\frac{\lambda}{\pi}}\sum_{i=1}^N V_i e^{-\lambda(x-x_i)^2}
V_y(y-y_i).
\end{equation}
This potential yields for $D_{p,l}$
\begin{eqnarray}
D_{p,l}=\overline{N_0} e^{-\frac{a^2}{8l_0 ^2}(l-p)^2}
\tilde{V_y} (a(l-p)/l_0^2)
\sum_{i=1}^N V_i
e^{i\frac{a}{l_0^2}(l-p)y_i}\nonumber\\
e^{-\frac{\overline{\lambda} }{2l_0^2}\left(x_i-\frac{a(l+p)}{2}\right)^2},
\end{eqnarray}
where  $\tilde{V_y}$ is the Fourier transform of $V_y$ ,
$\overline{N_0}=N_0\sqrt{\overline{\lambda}} $ and
$\overline{\lambda}=\lambda/(\lambda+B/2)$. $V_y$ could be any type of
potential, also
Gaussian.
All the above expressions are obtained without any approximations.
In the following, sensible
approximations that rely on experimental results will be introduced.
The systems we are considering are very pure samples in which the density
of impurities is
very low, essentially of the order of $10^{10}cm^{-2}$ \cite{koch1}. From
this it follows that the
average
distance between impurities is of the order of $d=1000\AA$. At this point it
is useful to note
that
in the lowest Landau level the typical magnetic length is of the order of
$l_0=100\AA$. Therefore,
the dimensionless quantity $d/l_0\simeq 10$, hence the limit we are
interested in is $d>>l_0$.
A natural assumption regarding the impurities is to
consider that they are all located on a super lattice of width $d$ where
$d/a$ is an integer.
The randomness of the impurities is therefore only given by the random
amplitudes
of the potentials. With these assumptions the positions of the impurities
can be redefined
with integer
coordinates on this lattice. Thus $V_{ij}$ is the impurity amplitude at the
position $(i,j)a$
and therefore in our system most $V_{ij}$ are zero.
In the usual
heterojunctions the elementary width is of the order of an $\AA$ngstr\"om
but as the relevant
length
scales of our system are essentially $l_0$ and $d$ we make the following
approximation: As we are
mainly concerned with the effect of the impurities, the elementary width
$a$ can be rescaled
to the same
order of magnitude as $d$ but with the condition $a<d$. Therefore the
typical bound needed to neglect the
terms $|p-l|>1$ in (7) is $a^2>8l_0^2$, which is equivalent to a nearest
neighbor approximation
and yields
\begin{eqnarray}
D_{l,l} & = & \overline{N_0}\tilde{V_y}(0) \sum_{i,j}^{N,M} V_{ij}
e^{-\frac{\overline{\lambda}a^2}{2l_0^2}\left(\frac{d}{a}i-l\right)^2}
\nonumber\\
D_{l,l+1} & = & \overline{N_0}\tilde{V_y}(a/l_0^2) e^{-\frac{a^2}{8l_0 ^2}}
\sum_{i,j}^{N,M} V_{ij} e^{i\frac{ad}{l_0^2}j}
\nonumber\\& & e^{-\frac{\overline{\lambda} a^2}{2l_0^2}\left(
\frac{d}{a}i-(l+1/2)\right)^2}.
\end{eqnarray}
We are now left with two types of terms: The diagonal terms $W_l
\equiv D_{l,l}$ and the
off-diagonal terms $D_{l+1}\equiv D_{l,l+1}$. With these new definitions
and using the fact that $d/a$ is
an integer, (for simplicity we will only consider even integers), we
obtain the following expressions
for equations (8)
\begin{equation}
W_{\frac{d}{a}l} = \overline{N_0}\tilde{V_y}(0) \sum_j^M V_{l,j}\quad,
\end{equation}
\begin{equation}
W_{\frac{d}{a}l+1} = W_{\frac{d}{a}l}\cdot
e^{-\frac{\overline{\lambda}a^2}{2l_0^2}} \mbox{ } ,\mbox{ }
W_{\frac{d}{a}l+2}= W_{\frac{d}{a}l}\cdot
e^{-4\frac{\overline{\lambda}a^2}{2l_0^2}} \mbox{ },\mbox{ } \cdots
\end{equation}
and for the off-diagonal terms we have
\begin{equation}
D_{\frac{d}{a}l+1} = D_{\frac{d}{a}l}\quad,\quad
D_{\frac{d}{a}l+2} = D_{\frac{d}{a}l-1}\quad,\quad\cdots
\end{equation}
These coefficients can now be inserted back in equation (5) yielding
\begin{equation}
(W_l-\epsilon)\psi_l+D_{l+1}\psi_{l+1}+D_l^*\psi_{l+1}=0.
\end{equation}
The usual procedure of transforming an off-diagonal disordered Anderson
model
to a diagonal disordered model \cite{flores} can now be applied, namely
\begin{equation}
\Psi_l=\psi_l/\phi_l \quad\mbox{where}\quad \phi_l^*\phi_{l-1}=1/D_l^* .
\end{equation}
The additional diagonal part simplifies to $1$ as $|\phi_l|=1$ due to (11).
The condition
that $d/a$ is an
even integer and a periodic boundary condition on the impurities is also
used. The transformed
equation (12) now
reads
\begin{equation}
(W_l-\epsilon)\Psi_l+\Psi_{l+1}+\Psi_{l-1}=0.
\end{equation}
If $\overline{\lambda}a^2>>2l_0^2$ we can neglect the terms (10).
 This states that only the
coefficients in the vicinity of an impurity survive, so that
 $W_l$ is non-zero if $l$ is a multiple of $d/a$, i.e.,
 $W_{\frac{d}{a}l}$ are random and $W_{\frac{d}{a}l+1}=W_{\frac{d}{a}l+2}=
 \cdots=0$

It is straightforward to see that the states at $\epsilon=0$ are {\em
overall extended},
as one only needs to suppose that $\Psi_{\frac{d}{a}l}=0$ and one is left
with an ordered
Anderson model.
The next step consists in studying the properties of the model around
$\epsilon=0$. For this
purpose we renormalize (14) in the following way
\begin{equation}
(W_{2l+1}-\epsilon)\Psi_{2l-2}+\Omega_{2l}(\epsilon)\Psi_{2l}+(W_{2l-1}-
\epsilon)\Psi_{2l+2}=0,
\end{equation}
where $\Omega_{2l}(\epsilon)=W_{2l+1}+W_{2l-1}-2\epsilon-(W_{2l+1}-
\epsilon)(W_{2l}-\epsilon)
(W_{2l-1}-\epsilon)$. Furthermore, in our diluted model $W_{2l+1}=0$,
which when
inserted in (15), yields
\begin{equation}
\Psi_{2l-2}+\left(2+\epsilon (W_{2l}-\epsilon )\right)\Psi_{2l}+
\Psi_{2l+2}=0,
\end{equation}
for $\epsilon\neq 0$. This last model was extensively studied in the
limit $\epsilon<<1$ by
Derrida and Gardner \cite{derrida}. They calculated the
complex Lyapounov exponent $\gamma$, where the real part corresponds to the
inverse localization
length and the imaginary part to $\pi$ times the integrated density of
states. Their results
can be expressed as follows:
\begin{equation}
\left\{\begin{array}{l}
Re(\gamma) \simeq K_1\epsilon^{2/3}\langle W^2\rangle ^{1/3}\nonumber\\
Im(\gamma) \simeq K_2\epsilon^{2/3}\langle W^2\rangle ^{1/3},
\end{array}\right.
\end{equation}
where $K_1=0.29\dots $ and $K_2=0.16\dots $ and $\langle\cdot\rangle$ is
the average over all
impurities.
{}From (17) it is straightforward that
\begin{equation}
\xi\sim\frac{1}{(E-E_0)^{2/3}\langle W^2\rangle ^{1/3}},
\end{equation}
and the density of states is
\begin{equation}
\rho(\epsilon)=\partial_{\epsilon}Im\gamma(\epsilon)\sim\epsilon^{-1/3}.
\end{equation}

Besides the assumption that all impurities are located on a square lattice,
we supposed that the
concentration of impurities is very low. Indeed,
Koch {\em et al.} \cite{koch1} measured the quantum Hall effect for different
types of scatterers. They
used $AlGaAs/GaAs$ heterostructures doped with either $Be$ $\delta$-like
repulsive scatterers or
$Si$ $\delta$-like attractive scatterers. They found that the exponent
$\kappa$ varied from
$0.36$ to
$0.81$ depending on the concentration of impurities, ranging from $0$ to
$2\cdot10^{10}cm^{-2}$,
 which corresponds to a typical distance between impurities of $d=1000\AA$.
  This is an indication that the impurity concentration might have an effect
   on the
localization length exponent and drives the system into a different
universality class,
depending on the concentration. As we
mentioned before the typical experimental value for $d/l_0$ is $10$ and
in our approximations
we used the fact that $d>a>>l_0$.
Nevertheless,
if we keep the first term of eq. (10), which corresponds to a second order
approximation,
we observe numerically no
change of the scaling (18) and (19). This shows that we don't have a
singular behavior due to the
first order approximation we used in our calculations and that our results
remain valid for less
extreme values of $d/l_0$.

Concerning the range of the potential and in order to remain
consistent within our approximations, the width
of the potential must be smaller or of the order of the magnetic length.
Actually in the limit
$\lambda\rightarrow
\infty$ ($\delta$-limit), Gredeskul {\em et al.} \cite{avishai} previously
demonstrated the
existence of extended
states exist in these
systems. We confirm their results and generalize them to non-$\delta$
potentials and evaluate the
scaling around the critical energy. Indeed in the edge state picture
\cite{halperin} of the
Quantum Hall effect
 one needs either the range of the potential to be larger than the
magnetic length, or $d>l_0$ if the range is shorter, when contacts are
taken into account
\cite{buttiker}. As this last condition applies, the model
studied fits in well with this picture to reproduce the features of the
Quantum Hall effect. For the
other approaches to the Quantum Hall effect, such as the original one
due to Laughlin \cite{laughlin},
the microscopic details of the impurities are unimportant, but states
must be extended at the
Landau levels and localized elsewhere, what we clearly demonstrated in
this letter.

A striking point in our results is that we obtain a different exponent
than most numerical studies.
The comparison with experiments is less trivial as their results have
still to be better
understood. The numerical studies, using finite size scaling \cite{hucke}
\cite{liu}, consider
the distance $d$ between impurities to be smaller, or of the same order
than the magnetic length
$l_0$ and more important,
no extended states exist, even at the Landau level; but the localization
length diverges when the
width tends to infinity. The different studies are in a good agreement
for a unique value
of the exponent $\nu\simeq2.35$, so that it has been conjectured that
they belong to a single
universality class. The case studied in this letter deals with the
opposite limit, i.e.,
$d>>l_0$ and we find a very different value for the exponent, namely
$\nu=2/3$. It is in fact not so surprising
that this model belongs to a different universality class as there is
a fundamental difference;
at the Landau level we have an overall extended state, even if the system
size is finite, so that
the transition from an extended state to a localized one is much sharper,
which in turn leads to a smaller
exponent. It would however be interesting to study the crossover between
both exponents, by
varying the impurity concentration.

It is worthwhile noticing that if we take the random potential to be zero,
this would imply that
$\epsilon=0$ in eq. (5) and that $\psi_l$ is
undefined, which reflects the degeneracy of a Landau level, so that the
density of states is
a $\delta$-function centered at the Landau level. Therefore the effect of
disorder is to broaden
the density
of states as $\epsilon^{-1/3}\langle W^2\rangle^{1/3}$, whereby only states
in the vicinity of $\epsilon=0$ remain
extended. For if
we equate the length of the system with the localization
length, so that the number of extended states, states whose localization
length exceeds
the size of the system, is independent of the size of the system,
but the transmission coefficient of eq. (14)
is one at $\epsilon=0$ \cite{m}.

The main result of this letter is the derivation of the localization length
 exponent $\nu$ analytically in
a two-dimensional disordered electron gas
in a high
external magnetic field. This exponent is found to be $2/3$. It is
calculated for short-ranged
impurity potentials when the concentration of impurities is low, i.e.,
 when the average distance
between impurities exceeds the magnetic length. Due to the low concentration
 of impurities, the system
belongs to a different universality class than most models studied
previously, which are closer to
a value of $\nu$ of around $2.3$. Additionally in the model studied
in this letter, states remain
overall extended at the center of the
Landau level, even if the width of the system is not infinite as opposed
to the
models studied numerically; and finally the density of states exhibits
the usual disorder
broadening.

It is a great pleasure to thank J.C. Flores for raising intriguing facts on
the subject and C.P. Enz and R. Ambigapathy for critical suggestions.
This work was supported in part by the Swiss National Science Foundation.

\end{document}